\documentclass[a4paper]{article}
\usepackage[psamsfonts]{amssymb}
\usepackage{amsmath}
\usepackage{epsfig}
\title{Quintom Cosmology with General Potentials\vspace{1cm}}
\author{\normalsize{M. R. Setare$^{1}$\thanks{%
E-mail: rezakord@ipm.ir}  \, and \,E.~N.~Saridakis $^{2}$\thanks{%
E-mail: msaridak@phys.uoa.gr} }\\
\newline
\\
{\normalsize \it $^1$ Department of Science, Payame Noor University,
Bijar, Iran}
\\
{\normalsize \it $^2$ Department of Physics, University of Athens,
GR-15771 Athens, Greece}
\\
}
\date{}
\begin{document}
\maketitle
\newcommand{\be}{\begin{equation}}
\newcommand{\ee}{\end{equation}}
\newcommand{\bq}{\begin{eqnarray}}
\newcommand{\eq}{\end{eqnarray}}
\vspace{1cm}
 \abstract
We investigate the phase-space structure of the quintom paradigm
in the framework of a spatially flat, open, or closed isotropic
and homogeneous universe. We examine the dynamical evolution under
the assumption of late-time dark energy domination, without
specifying the explicit quintom potential form. The obtained
cosmological behavior is qualitatively different than that
acquired from the single phantom model.

\vspace{1cm}
\newpage
\section{Introduction}

One of the most important problems of cosmology, is that of dark
energy. Type Ia supernova observations suggest that the universe
is dominated by dark energy with negative pressure, which provides
the dynamical mechanism for the accelerating expansion of the
universe \cite{{per},{gar},{ries}}. The strength of this
acceleration is presently a matter of debate, mainly because it
depends on the theoretical model implied when interpreting the
data. Most of these paradigms are based on the dynamics of a
scalar or multi-scalar fields (e.g quintessence
\cite{{rat},{zlat}} and quintom \cite{{quintom},{quint2}} models
of dark energy, respectively). In addition, other proposals on
dark energy include interacting dark energy models \cite{intde},
braneworld models \cite{brane}, Chaplygin gas models \cite{cg},
holographic dark energy \cite{holoext}, bulk holographic dark
energy \cite{bulkhol} and many others.

Initially, a scalar field candidate for dark energy was the
quintessence scenario, which is based on a fluid with
equation-of-state parameter lying in the range, $-1< w< {-1/3}$.
However, although investigations using many other models lead also
to universe-acceleration below the de Sitter value \cite{Daly}, it
is certainly true that the body of observational data allows for a
wide parameter space compatible with an acceleration larger than
this divide \cite{{cald},{hans}}. If eventually this turn out to
be the case, then the fluid driving the expansion would violate
not only the strong energy condition $\rho + 3P>0$, but the
dominant energy condition $\rho + P>0$, too. Fluids (fields) of
such behavior are dubbed phantom fluids \cite{caldwell}. In spite
of the fact that the theory of phantom fields encounter the
problem of stability, which one could try to bypass by assuming
them to be effective fields \cite{{car},{gib}}, it is nevertheless
interesting to study their cosmological implications and there are
many relevant studies recently on phantom energy \cite{meng}.

Although observations mildly favor dark energy models where $w$
has crossed -1 in the near past, neither quintessence nor phantom
can fulfill this transition. In quintessence model the equation of
state $w=p/\rho$ is always in the range $-1\leq w\leq 1$ for
$V(\phi)>0$. Meanwhile, in the phantom scenario, which comparing
to the quintessence has the opposite sign in the kinetic term in
the Lagrangian, one always obtains $w\leq -1$. Therefore, neither
of these two models alone can fulfill the transition from $w>-1$
to $w<-1$ and vice versa. Furthermore, although in k-essence
\cite{k-essence} one can have both $w\ge -1$ and $w<-1$, it has
been lately shown in \cite{Vikman1, Vikman2} that the
corresponding crossing is very unlikely to be realized during the
evolution. However, one can show \cite{{quintom},{quint2}} that
considering the combination of quintessence and phantom in a
qualitatively new model, the $-1$-transition can be fulfilled, as
can be clearly seen in \cite{ref}. This model, dubbed quintom, can
produce a better fit to observational data than the more
conventional paradigms. Finally, note that in the recent work
\cite{Xia} the authors prove in full generality the no-go theorem
that in order to acquire the $w=-1$ crossing it is necessary to
have more than one degrees of freedom. This theorem offers a
concrete theoretical justification for the quintom paradigm.

In the literature, there has been a general investigation of the
phase-space of a spatially flat homogeneous and isotropic universe
dominated by a phantom field. In \cite{far}, the peculiar dynamics
arising from a negative kinetic energy density has been studied
with the help of a toy model consisting of two coupled
oscillators, one with negative kinetic energy representing the
phantom field $\phi$, and the other with positive kinetic energy
mimicking the gravitational field. The spacetime geometry is that
of a spatially flat Friedmann-Lemaitre-Robertson-Walker (FLRW)
universe, the assumed potentials maintain a general form, and the
asymptotic dynamics (late-time attractors) is examined in the
three-dimensional phase-space $(H,\phi,\dot{\phi})$, under the
condition of the final domination of the phantom. On the other
hand, in \cite{hao} the authors have examined several scenarios
which exhibit a late-time behavior where cold matter with zero
pressure dominates the evolution and the phantom energy decays.

In the present study we desire to perform the phase-space
investigation in the case of quintom cosmology, constructed using
both phantom ($\phi$) and quintessence ($\sigma$) fields. As was
mentioned above, despite its outward structural similarity,
quintom behavior is conceptually and qualitatively different from
the pantom one. We desire to maintain a non-specific potential
form $V(\phi,\sigma)$, and we assume that quintom dark energy
eventually dominates the cosmic dynamics. Considering a
non-spatially flat FLRW universe as the underlying spacetime
geometry, we study quintom dynamics and we examine whether the
universe decomposes in a singularity or expands forever.

\section{Quintom cosmology}
\label{Quintom cosmology}

We are interesting in investigating the late-time cosmological
evolution of the quintom field, consisting of the normal scalar
field $\sigma$ and the negative-kinetic-energy scalar field
$\phi$. We consider a general Friedmann-Lemaitre-Robertson-Walker
universe with line element
 \be\label{metr}
 ds^{2}=-dt^{2}+a^{2}(t)\left(\frac{dr^2}{1-kr^2}+r^2d\Omega^{2}\right)
 \ee
 in comoving coordinates  $(t,r,\theta,\varphi)$, where $a$ is the scale factor and $k$ denotes
 the space curvature with $k=0,1,-1$ corresponding to a flat, closed
or open universe respectively.

 The action which describes the
quintom model is expressed in the following form
\cite{val,setsar}:
\begin{equation}
\label{1} S=\int d^4x\sqrt{-g}\left[\frac{R}{2}
 -\frac{1}{2}g^{\mu \nu}\partial _\mu \phi \partial _\nu \phi
 +\frac{1}{2}g^{\mu \nu}\partial _\mu \sigma \partial _\nu \sigma
 +V(\phi ,\sigma)\right],
\end{equation}
where we have neglected the Lagrangian density of matter fields
consistently to their late-time downgrading.
 The effective energy density $\rho$ and the effective
pressure $P$ of the scalar fields, are given by \cite{quintom}:
 \begin{equation}
 \rho=-\frac{1}{2}\dot{\phi}^2+\frac{1}{2}\dot{\sigma}^2+V(\phi,\sigma)
 \label{rho}
\end{equation}
\begin{equation}
P=-\frac{1}{2}\dot{\phi}^2+\frac{1}{2}\dot{\sigma}^2-V(\phi,\sigma),
 \label{Pe}
\end{equation}
where the negative sign of $\phi$-kinetic energy is characteristic
of a phantom field.

The first Einstein field equation is:
\begin{equation}
 H^{2}+\frac{k}{a^{2}}=\frac{8\pi}{3M_p^{2}}\rho,
\label{eq1}
 \end{equation}
 where $H$ is the Hubble parameter defined, as usual, as
 $H\equiv\dot{a}/a$.
 Equation (\ref{eq1}) can be considered as a constraint
for the Hubble rate. Setting $8\pi/M_p^{2}=1$ and using
(\ref{rho}), it is rewritten as:
\begin{equation}
H^{2}+\frac{k}{a^{2}}=\frac{1}{6}\left[-\dot{\phi}^{2}
+\dot{\sigma}^{2}+2V(\phi,\sigma)\right]. \label{friedmann}
\end{equation}
The other FLRW equation is:
\begin{equation}
\dot{H}=\frac{k}{a^{2}}+\frac{1}{2}\left(\dot{\phi}^{2}-\dot{\sigma^{2}}\right).
\label{Hdot}
\end{equation}
Furthermore, the evolution equations for the two scalar fields in
FLRW framework have the following form:
\begin{equation}
\ddot{\phi}+3H\dot{\phi}-\frac{\partial V}{\partial \phi}=0
 \label{phieq}
\end{equation}
\begin{equation}
\ddot{\sigma}+3H\dot{\sigma}+\frac{\partial V}{\partial \sigma}=0,
 \label{sigmaeq}
\end{equation}
where dots denote differentiation with respect to the comoving
time $t$. Note also that, as usual, only three of equations
(\ref{friedmann})-(\ref{sigmaeq}) are independent. Lastly, we
mention that the solutions of (\ref{friedmann})-(\ref{sigmaeq})
must lead to a non-negative total quintom energy density (given in
(\ref{rho})).

Finally, it will be helpful to notice that alternatively we can
derive the field equations form the Lagrangian:
\begin{equation}
\mathbf{L}=a^{3}(\rho-P)=3a\dot{a}^{2}-3ak+a^{3}\left[\frac{1}{2}\dot{\phi}^{2}-\frac{1}{2}\dot{\sigma}^{2}+V(\phi,\sigma)\right]
 \label{Lagr}
\end{equation}
or, equivalently from the Hamiltonian:
\begin{equation}
\mathbf{H}=3a^{3}\left[H^{2}-\frac{k}{a^{2}}+\frac{1}{6}\left(\dot{\phi}^{2}-\dot{\sigma}^{2}\right)-\frac{1}{3}V(\phi,\sigma)\right].
 \label{Hami}
\end{equation}

\section{The phase-space of the quintom paradigm}
\label{phase space}

In the previous section we formulated the dynamical equations for
the quintom model. In the present section we are interested in
investigating its phase-space behavior for the various curvature
cases.

It is obvious that in a spatially-flat FLRW universe
(\ref{friedmann}) corresponds to $\mathbf{H}=0$. Thus, for $k=0$
equation (\ref{friedmann}) constitutes a Hamiltonian constraint
which in the phase-space confines the orbits of the dynamical
solutions of equations (\ref{friedmann})-(\ref{sigmaeq}) to the
surface of constant energy $\mathbf{H}=0$ \cite{far}. However, for
$k\neq0$ we obtain $\mathbf{H}\neq0$ and in particular
$\mathbf{H}=-6ak$. Thus, in a spatially non-flat universe this
relation provides at each time a surface of constant energy in the
phase-space, for orbits of the solutions of equations
(\ref{friedmann})-(\ref{sigmaeq}), and this surface is in general
curved. In the following, instead of the scale factor, we choose
the Hubble parameter to be a dynamical variable (since this is
indeed the observable quantity) and thus the phase-space is
five-dimensional: $(H,\phi,\dot{\phi},\sigma,\dot{\sigma})$.

When the values of $H$, $\phi$, $\sigma$ and $\dot{\sigma}$ are
specified, we can exert (\ref{friedmann}) in order to deduce the
values of $\dot{\phi}$:
\begin{equation}
\dot{\phi}^{2}=\dot{\sigma}^{2}+2V-6\left(H^{2}+\frac{k}{a^{2}}\right).
\label{phidot}
\end{equation}
This relation determines the sub-space of the phase-space that is
forbidden for the dynamics. Indeed, defining the quantity
\begin{equation}Q=\dot{\sigma}^{2}+2V-6\left(H^{2}+k/a^{2}\right),\end{equation}
(\ref{phidot}) implies that the forbidden sub-space $\mathcal{F}$
is specified as:
\begin{equation}
\mathcal{F}\equiv\{(H,\phi,\dot{\phi},\sigma,\dot{\sigma}): \ \ \
\ Q<0\}.
\end{equation}
Furthermore, in order to examine the properties of the allowed
sub-space, we deduce from (\ref{phidot}) that:
\begin{equation}
\dot{\phi}=\pm\sqrt{Q}. \label{phidot2}
\end{equation}
Thus, for every $H$, $\phi$, $\sigma$ and $\dot{\sigma}$, there
are two distinct values of $\dot{\phi}$. Therefore, we conclude
that the sub-space of the five-dimensional phase-space that is
accessible by the dynamics, consists of two four-dimensional
sub-surfaces, one for each solution branch of (\ref{phidot2}). We
call sub-surface $\mathcal{A}$ the one corresponding to the
positive branch, and $\mathcal{B}$ the one arising from the
negative branch. These two sub-surfaces meet at the
three-dimensional sub-surface $\mathcal{C}$, which is defined as:
\begin{equation}
\mathcal{C}\equiv\left\{(H,\phi,\dot{\phi},\sigma,\dot{\sigma}):\
\ \ \
V(\phi,\sigma)=3\left(H^{2}+\frac{k}{a^{2}}\right)-\frac{\dot{\sigma}^2}{2},
\ \  \dot{\phi}=0\right\},\label{CC}
\end{equation}
which belongs also to the boundary of the forbidden region
$\mathcal{F}$ and lies in the $(H,\phi,\sigma,\dot{\sigma})$
plane.

Let us now examine the fixed points of the system
(\ref{friedmann})-(\ref{sigmaeq}), which as usual are quested
using the conditions: $\dot{\phi}=0$, $\dot{\sigma}=0$ and
$\dot{H}=0$. For the flat universe case ($k=0$) these points
$(H_0,\phi_0,\sigma_0)$ are easily extracted and they are
determined by:
\begin{equation}
H_0=\pm\sqrt{\frac{V_0}{3}} ,   \ \ \ \ \frac{\partial V}{\partial
\phi}|_0=0, \ \ \ \ \frac{\partial V}{\partial
\sigma}|_0=0,\label{fixed0}
\end{equation}
where $V_0\equiv V(\phi_0,\sigma_0)$ and the derivatives are
calculated at this potential point. Thus, for the flat universe
the equilibrium points for the dynamics are simply the de Sitter
spaces with constant scalar fields. Note however than in order for
these points to exist, the corresponding cosmological potentials
must have points of zero gradient in both $\phi$ and $\sigma$
directions. Furthermore, from (\ref{CC}) and (\ref{fixed0}) we
deduce that the fixed points lie on the sub-space $\mathcal{C}$.

Let us make here a comment on the two distinct solutions of
relation (\ref{fixed0}). As it was shown in \cite{Cai06}, there is
a cosmic duality between different quintom solutions. In
particular, an expanding universe, which at early times is
dominated by quintessence ($w>-1$) and lately by phantom ($w<-1$)
fields, is dual to a contracting universe which faces the
transition from $w<-1$ to $w>-1$. It is easy to see that the two
distinct fixed-points given in (\ref{fixed0}), correspond to such
expanding and contracting universes. This property is helpful for
the investigation of the corresponding cosmological evolution.

For the $k\neq0$ case the dynamical system
(\ref{friedmann})-(\ref{sigmaeq}) does not have fixed points,
apart from the two extreme solutions, one with $H=0$,
$a\rightarrow\infty$,  $\dot{a}<\infty$, $V_0=0$, and $V$'s
derivatives equal to zero, and the other with
$a\rightarrow\infty$,  $\dot{a}\rightarrow\infty$,
$H=\pm\sqrt{V_0/3}$, and $V$'s derivatives equal to zero. Thus, in
the non-flat universe our model is always out of equilibrium. In
this case it is useful to take advantage of the aforementioned
duality between the different quintom solutions \cite{Cai06}.
Namely, as we can easily see, an eternally expanding universe
starting with an initial singularity at $t=0^+$ is dual to a
contracting one that begins with an infinite scale factor. In
addition, a contracting universe ending in a Big Crunch at
$t=0^-$, is dual to an expanding universe ending in a final Big
Rip at $t=0^-$.  This behavior is definitely qualitatively
different from the flat paradigm. Obviously, the realization of a
particular branch depends on the initial conditions.

It would be interesting to examine the case of an orbit with
initial conditions chosen exactly on $\mathcal{C}$. We generalize
\cite{far} and we construct the tangent to the orbits as:
\begin{equation}
\overrightarrow{T}\equiv(\dot{H},\dot{\phi},\ddot{\phi},\dot{\sigma},\ddot{\sigma})
\equiv\left(\frac{1}{2}\dot{\phi}^{2}-\frac{1}{2}\dot{\sigma}^{2},\dot{\phi},
\frac{\partial V}{\partial
\phi}-3H\dot{\phi},\dot{\sigma},-\frac{\partial V}{\partial
\sigma}-3H\dot{\sigma}\right),
\end{equation}
in the $(H,\phi,\dot{\phi},\sigma,\dot{\sigma})$ space. On the
boundary $\mathcal{C}$ we have $\dot{\phi}=0$ and thus the
subsequent dynamics will depend on $\dot{\sigma}$ and on the
potential derivatives. If additionally $\dot{\sigma}=0$, then $
\overrightarrow{T_{\mathcal{C}}}=(0,0,\frac{\partial V}{\partial
\phi},0,-\frac{\partial V}{\partial \sigma})$. Therefore, if
$\partial V/\partial \phi|_{\mathcal{C}}>0$ and $\partial
V/\partial \sigma|_{\mathcal{C}}<0$, an orbit beginning on
$\mathcal{C}$ will move into sub-surface $\mathcal{A}$, whereas if
$\partial V/\partial \phi|_{\mathcal{C}}<0$ and $\partial
V/\partial \sigma|_{\mathcal{C}}>0$ it will move inside
sub-surface $\mathcal{B}$. For the crossed cases, we can not
predict the specific evolution unless we know the explicit
potential gradients. Note that these results hold for both flat
and non-flat universe, in a unified way.

Finally, we are interesting in investigating the evolution of the
Hubble parameter. In the case of a single phantom field in a flat
universe background, it was shown in \cite{far} that $\dot{H}>0$
everywhere apart from the fixed points, and this behavior
corresponds to super-acceleration \cite{fara}. However, in the
present quintom model, $H$-behavior is different. Indeed, for all
$k$ cases the sign of $\dot{H}$, given by (\ref{Hdot}), can be
varying and moreover (according to the values of $\dot{\phi}$ and
$\dot{\sigma}$ at each moment) it can vary between the different
stages of a specific universe evolution. Moreover, since $H$ is
not evolving monotonically, periodic orbits cannot be excluded a
priori, and indeed there could be potentials leading to such a
behavior, or even to cyclic universes \cite{cyclic}.

Although in the aforementioned analysis, in order to examine the
general evolution characteristics, we have remained in the general
potential case,  let us finish this section by giving a specific
example using  a simple potential form. We consider the quadratic
potential
$V(\phi,\sigma)=\frac{1}{2}m^2_\phi\phi^2+\frac{1}{2}m^2_\sigma\sigma^2+c_p$,
where $m_\phi$ is the quintessence component mass, $m_\sigma$ is
the phantom component mass and $c_p$ is a constant. In this case,
for the flat universe case the corresponding fixed-points are the
loci where $\phi=\sigma=0$ and $H_0=\pm\sqrt{c_p/3}$. Thus, for
$c_p\neq0$ we obtain de Sitter universes, while for $c_p=0$ we
obtain Minkowski spaces. Knowing the specific potential form it is
straightforward to perform a stability analysis and show that
these fixed-points are unstable. These results are in agreement
with \cite{Cai06,setsar}. Note also the duality between expanding
and contracting universes (the two distinct $H_0$ cases) described
above. In addition, for this quadratic case and imposing
$\dot{\sigma}=0$, we obtain
$\overrightarrow{T_{\mathcal{C}}}=(0,0,m^2_\phi\phi^2,0,-m^2_\sigma\sigma^2)$,
leading to the result that an orbit beginning on $\mathcal{C}$
will move into sub-surface $\mathcal{A}$.

\section{Conclusions}
\label{Conclusions}

In this work we have investigated the phase-space structure of the
quintom paradigm without determining a specific potential form.
Furthermore, our analysis was performed in a general FLRW
geometrical background, covering the cases of a flat, open and
closed universe. The quintom scenario is realized by a dynamical
system consisting of two scalar fields, one with negative and the
other with positive kinetic energy, where we assume that at late
times the quintom fields dominate the dynamics. Our results are
qualitatively different comparing to the single phantom model
examined in \cite{far}, thus offering an additional argument for
the conceptual difference between quintom and phantom cosmology.

The phase-space of the quintom model proves to consist of two
connected four-dimensional sub-surfaces in the
$(H,\phi,\dot{\phi},\sigma,\dot{\sigma})$ space. In the flat
universe framework, and provided that the potential
$V(\phi,\sigma)$ possesses points of zero gradient in both $\phi$
and $\sigma$ directions, the equilibrium solutions of the dynamics
are just de Sitter spaces. In such cases the expansion may last
for ever. However, in non-flat universes the behavior can be
qualitatively different, since the dynamical motion is at all
times out of equilibrium, and the universe is expanding forever or
it results to a Big Rip.

Our investigation reveals that the Hubble parameter is a
non-monotonic function of time, and this is a radical difference
comparing to the single phantom model \cite{far}. Indeed, $H$ can
be increasing or decreasing in subsequent evolution stages of a
specific universe. Furthermore, there could be special potential
cases leading to periodic behavior, i.e to cyclic universes.

Quintom scenario offers an efficient explanation of the $-1$
crossing of the equation-of-state parameter of dark energy, thus
being consistent with observations. From our investigations it is
implied that it additionally offers a significantly larger variety
of cosmological evolutions than the simple phantom model. These
characteristics make quintom cosmology an interesting subject for
further investigation.

\vskip .1in \noindent {\bf {\large Acknowledgment}}

The authors would like to thank V.~Faraoni for useful discussions
and an anonymous referee for helpful comments.

\end{document}